\documentclass[paper]{JHEP3}
\usepackage{amsmath,amssymb}

\newcommand{\TeXmacs}{T\kern-.1667em\lower.5ex\hbox{E}\kern-.125emX\kern-.1em\lower.5ex\hbox{\textsc{m\kern-.05ema\kern-.125emc\kern-.05ems}}}
\newcommand{\tmem}[1]{{\em #1\/}}
\newcommand{\tmop}[1]{\ensuremath{\operatorname{#1}}}
\newcommand{\tmtextit}[1]{{\itshape{#1}}}
\newcommand{\tmtextsf}[1]{{\sffamily{#1}}}
\newcommand{\tmtexttt}[1]{{\ttfamily{#1}}}
\newcommand{\withTeXmacstext}{This document has been produced using \TeXmacs (see \texttt{http://www.texmacs.org})}
\newenvironment{itemizedot}{\begin{itemize} }{\end{itemize}}

\preprint{Bicocca-FT-07-13}

\title{MINT: a Computer Program for Adaptive Monte Carlo Integration and
    Generation of Unweighted
    Distributions\thanks{{\withTeXmacstext}}}
\author{Paolo Nason\\
    INFN, Sezione di Milano-Bicocca, Piazza della Scienza 3, 20126 Milan,
    Italy\\
 E-mail: \email{Paolo.Nason@mib.infn.it}}

\abstract{
  In this note I illustrate the program \tmtexttt{MINT}, a FORTRAN program for
  Monte Carlo adaptive integration and generation of unweighted distributions.}

\begin{document}

\section{Introduction}

The purpose of this note is to illustrate the program \tmtexttt{MINT}, a
FORTRAN program for Monte Carlo adaptive integration and generation of
unweighted distributions. The program performs the following task: given a
function $f (x^1, \ldots, x^n)$, defined in the unit $n$-dimensional cube, and
such that the function
\begin{equation}
  \tilde{f} (x^1, \ldots, x^p) = \int_0^1 d x^{p + 1} \int_0^1 d x^{p + 2}
  \ldots \int_0^1 d x^n f (x^1, \ldots, x^n) \label{barred}
\end{equation}
is positive for $x^1, \ldots, x^p$ in the $p$-dimensional unit cube, it
computes the integral of $f$, and generates $x^1, \ldots, x^p$ points in the
unit cube, distributed with probability
\begin{equation}
  \tilde{f} (x^1, \ldots, x^p) d x^1 \ldots d x^p .
\end{equation}
The case $p = n$ is contemplated, i.e. \tmtexttt{MINT} can also perform the
task of unweighting a positive distribution.

A popular program to perform adaptive Monte Carlo multi-dimensional
integration is the VEGAS program {\cite{Lepage:1980dq}}. It uses an
importance-sampling method, where the sampling rate is independent for each
coordinate. The method works well for factorizable singularities, and it has
the advantage that the sampling information one needs to store grows only
linearly with the number of dimensions.

A common problem encountered in particle physics phenomenology, besides the
integration of a given multidimensional function, is the generation of its
arguments with a probability proportional to its value. The SPRING-BASES
program {\cite{Kawabata:1995th}} use the VEGAS algorithm to perform the
integration of a positive function, and then can generate events distributed
with a probability proportional to the integrand. It first stores the
integration result and the maximum value of the function for each cell of the
adaptive mesh found by VEGAS. In the generation stage, a cell is chosen with a
probability proportional to the corresponding value of the integral, and then
a point in the cell is generated using the hit and miss technique. This method
is highly efficient, but it has the disadvantage that the amount of storage
space it requires grows exponentially with the dimension.

MINT is a replacement for the SPRING-BASES package. It differs from it in the
following way. In order to generate the phase-space point, it does not store
the value of the integral for each cell. It only stores an upper bound of the
value of the function in each cell. The multidimensional stepwise function
that equals the upper bound of the function to be integrated in each cell is
in fact an upper bound for the whole function, and it is easy to generate
phase space points distributed according to it. Using again the hit and miss
technique, we can then generate points according to the original distribution.
In order to save space, MINT uses an upper bounding functions that is the
products of a set of step-wise functions, each of them associated with a
coordinate. The storage requirement for such a function grows only linearly
with the dimension. The ability to deal with non-positive functions is
achieved in \tmtexttt{MINT} by {\tmem{folding over}} (in a sense that will be
specified later) the integrand along the directions $p + 1, \ldots, n$.

The problem of unweighting a distribution of the form (\ref{barred}) arises in
the context of the method of refs.
{\cite{Nason:2004rx,Nason:2006hf,Frixione:2007nw}} for the inclusion of
next-to-leading corrections in Monte Carlo generated events.

\section{The algorithm}

The algorithm is in essence the VEGAS algorithm. It is implemented in
\tmtexttt{MINT} in the following way. Assume that we deal with the
$n$-dimensional integral of a function $f (x^1, \ldots, x^n) \geqslant 0$ in
the unit hypercube. We divide the $[0, 1]$ interval for each coordinate in $m$
bins of variable length,
\[ x_{l - 1}^k \leqslant x^k \leqslant x_l^k \tmop{for} l = 1, \ldots m
   \tmop{and} k = 1, \ldots, n . \]
We then define $n$ monotonic, continuous functions $h^k (y^k)$, with $0 < y^k
< 1$ and $k = 1, \ldots, n$, such that
\begin{equation}
  h^k \left( \frac{l}{m} \right) = x^k_l \tmop{for} l = 0, \ldots, m,
\end{equation}
and linear (i.e. having constant first derivative) in all the intervals $(l -
1) / m < y < l / m$.

We have
\begin{equation}
  \int f (x) d^n x = \int f \left( h (y) \right)  \prod^n_{k = 1} \frac{d h^k
  (y^k)}{d y^k} d^{} y^k .
\end{equation}
Observe that, given $y^k$, if $l$ is the bin where $y$ lies (i.e. $(l - 1) / m
< y < l / m$), we have
\begin{equation}
  \frac{d h^k (y^k)}{d y^k} = \left( x^k_l - x^k_{l - 1} \right) \times m .
\end{equation}

We begin the adaptation process with $x^k_l = l / m$ for all $k = 1, \ldots,
n$, $l = 1, \ldots, m$. We would like to find optimal $h^k$ functions, such
that the integration process has small errors. This is done as follows. We
perform several iteration of the integration. In each iteration we generate a
set $\mathcal{I}_N$ of $N$ random points $y$ in the unit hypercube $0
\leqslant y^k \leqslant 1$, $k = 1, \ldots ., n$, and estimate the integral
with the formula
\begin{equation}
  I = \frac{1}{N} \sum_{y \in \mathcal{I}_N} f (h (y)) \prod^n_{k = 1} \frac{d
  h^k (y^k)}{d y^k} .
\end{equation}
To each $y$ we associate a point $x = h (y)$, and we call $l^k$ the bin where
$x^k$ lies. We accumulate the value of $f$ and the number of hits in two
arrays $R^k_{^{} l}$ and $N^k_{l^{}}$, with $k = 1, \ldots, n$ and $l = 1,
\ldots, m$
\begin{equation}
  R^k_{^{} l} = \sum_{y \in \mathcal{I}_N} \theta (l - 1 \leqslant y^k < l)
  \times f (x) \prod_{k' = 1}^n (x^{k'}_{l^k} - x^{k'}_{l^k - 1}), N^k_l =
  \sum_{y \in \mathcal{I}_N} \theta (l - 1 \leqslant y^k < l),
\end{equation}
where
\[ \prod_{k = 1}^n (x^k_{l^k} - x^k_{l^k - 1}) \]
is the volume of the cell where $x$ lies. After all the $N$ points have been
generated, we define for all \ $k = 1, \ldots ., n$ \ and $l = 0, \ldots, m$
\begin{equation}
  I^k_l = \sum_{j = 1}^l \frac{R^k_j}{N^k_j}, I^k_0 = 0
\end{equation}
and a corresponding continuous piecewise linear function $i^k (x^k)$ such that
$i^k (x^k_l) = I^k_l$ for $l = 0, \ldots ., m$. Notice that $i^k ( \bar{x}^k)$
is an estimate of the integral of $f$ in the $n - 1$ dimensional
hyper-rectangle defined by the condition $0 \leqslant x^k \leqslant
\bar{x}^k$. We now find the new $x^k_l$ points by solving the equation
\begin{equation}
  i^k ( \text{$x^k_l$}) = i^k (1) \frac{l}{m}
\end{equation}
and then we go to the next iteration. It is clear that when all the $R^k_j /
N^k_j$ are equal, the procedure has reached stability, the $x^k_l$ no longer
change, and we are probing regions of equal importance with (in the average)
the same number of points. Furthermore, this procedure is nearly optimal if
\begin{equation}
  f (x^1, \ldots ., x^n) = f_1 (x^1) \times \ldots . \times f_n (x^n) .
\end{equation}

\section{Folded integration}

If the given function $f$ is not positive definite, but its integral over a
subset of the integration variables is positive, we can turn it into a
positive function by folding it over itself a sufficient number of times in
the given subset of the integration variables. More precisely, we do the
following. Suppose we want to fold each $k$ coordinate $p_k$ times. We define
the function
\begin{equation}
  \label{eq:foldedint} \bar{f} (z^1, \ldots, z^n) = \frac{1}{p_1}  \sum_{l_1 =
  0}^{p_1 - 1} \ldots \frac{1}{p_n}  \sum_{l_n = 0}^{p_n - 1} \left[ f \left(
  h (y) \right)  \prod^n_{k = 1} \frac{d h^k (y^k)}{d y^k} \right]_{y = y (z,
  l)},
\end{equation}
where
\[ y^k (z^k, l^k) = \frac{l_k + z^k}{p_k} \]
so that, as the variable $z^k$ spans the $[0, 1]$ interval, $y^k$ spans each
of the $p_k$ equal subintervals
\begin{equation}
  \frac{l_k}{p_k} < y^k < \frac{l_k + 1}{p_k} .
\end{equation}
It is clear that
\begin{equation}
  \int f (x) d^n x = \int f \left( h (y) \right)  \prod^n_{k = 1} \frac{d h^k
  (y^k)}{d y^k} d^{} y^k = \int \bar{f} \left( z^1, \ldots, z^n \right) d z^1
  \ldots d z^n .
\end{equation}
It is convenient for practical purposes to choose the $p_k$ among the integer
divisors of $m$.

\section{Generation of $p$-tuples $x^1, \ldots, x^q$.}

In order to generate $p$-tuples $x^1, \ldots, x^p$, with $p \leqslant n$, and
a probability distribution
\begin{equation}
  P (x^1, \ldots, x^p) = \left\{ \begin{array}{ll}
    \frac{\int f (x) d x^{p + 1} \ldots d x^n}{\int f (x) d^n x}  & \tmop{for}
    p < n\\
    \frac{f (x)}{\int f (x) d^n x} & \tmop{for} p = n
  \end{array}, \right.
\end{equation}
we generate the whole $n$-tuple, and then discard the $p + 1, \ldots, n$
coordinates. We can use an arbitrary amount of folding on the $p + 1, \ldots,
n$ coordinates, while we must keep the $1, \ldots, p$ coordinates unfolded
(i.e. $p^1, \ldots, p^q = 1$). We then proceed with the generation of the $z$
variables using the $\bar{f}$ function, with probability
\begin{equation}
  P (z^1, \ldots, z^n) = \frac{\bar{f} \left( z^1, \ldots, z^n \right)}{\int
  \bar{f} \left( z^1, \ldots, z^n \right) d z^1 \ldots d z^n} .
\end{equation}
We seek for $\bar{f}$ an upper bound of the form
\begin{equation}
  \bar{f} (z^1, \ldots, z^n) \leqslant u^1 (z_1) \times \ldots \times u^n
  (z_n),
\end{equation}
where $u^k (z^k)$ are stepwise functions in the $0 < z^k < 1$ interval divided
into $m / p_k$ equal subintervals. The $u^k (z^k)$ are initialized according
to
\begin{equation}
  u^k (z^k) = \left[ \int \bar{f} \left( z^1, \ldots, z^n \right) d z^1 \ldots
  d z^n \right]^{1 / n} .
\end{equation}
We perform a large number of calls to the function $\bar{f}$, at (uniform)
random values of its argument. If the bound is violated
\begin{equation}
  \bar{f} (z^1, \ldots, z^n) \geqslant u^1 (z_1) \times \ldots \times u^n
  (z_n),
\end{equation}
each of the $u^k (z_k)$ is increased by a fixed factor $f$ in the subinterval
containing $z_k$. The value
\begin{equation}
  f = 1 + \frac{1}{10 n}
\end{equation}
is found to work reasonably well. After a sufficiently large number of calls,
the $u$ will stabilize.

In practice, in the MINT program, the upper bounding envelope is computed
during the folded integration.

\section{The code}

The code is available at the URL
\begin{center}
{\tt http://moby.mib.infn.it/\~{}nason/POWHEG/FNOpaper/mint-integrator.f}\ .
\end{center}
It is composed by two user-callable routines, \tmtextsf{\tmtexttt{mint}}
and \tmtextsf{\tmtexttt{gen}}. Furthermore, the common block\\
\tmtexttt{ \ \ \ \ \ integer ifold(ndimmax)\\
\ \ \ \ \ common/cifold/ifold}\\
must be set, specifying how many times each dimension is folded. The user
function must have the form\\
\tmtexttt{ \ \ \ \ \ function fun(x,w,ifl)\\
\ \ \ \ \ real * 8 fun,w,x(ndim)}\\
When called with \tmtexttt{ifl=0} it must return
\tmtextsf{\tmtexttt{fun=f(x)*w}}. When called with \tmtexttt{ifl=1} it returns
\tmtextsf{\tmtexttt{fun=f(x)*w}}, but it may assume that, since the previous
\tmtextsf{\tmtexttt{ifl=0}} call the variables that are not folded, and all
the values that have been computed with them, have remain unchanged. The
return value \tmtexttt{f(x)*w} is \tmtextit{not} used by \tmtexttt{mint} in
this case. When called with \tmtexttt{ifl=2}, \tmtexttt{fun} must return the
sum of all the return values up to (and including) the last \tmtexttt{ifl=0}
call. This apparently cumbersome procedure is needed to allow for enough
flexibility, in order for a program to be able to compute the fraction of
positive and negative contributions to the folded integral, which is needed in
POWHEG applications.

To run the program, one first calls the subroutine
\begin{verbatim}
      real * 8 xgrid(50,ndimmax),xint,ymax(50,ndimmax),ans,err
      integer ndim,ncalls0,nitmax,imode
      call mint(fun,ndim,ncalls0,nitmax,imode,xgrid,xint,ymax,ans,err)
c ndim: dimension of x in fun (ndim<=ndimmax)
c ncalls0: maximum number of calls per iteration
c nitmax: number of iterations
c imode: flag
\end{verbatim}
\begin{itemizedot}
  \item When called with \tmtexttt{imode=0}, \tmtexttt{mint} performs the
  integration of the absolute value of the function, finds the optimal grid
  \tmtexttt{xgrid}, stores the answer in \tmtexttt{xint} and \tmtexttt{ans},
  and the error in \tmtexttt{err}.
  
  \item When called with \tmtexttt{imode=1}, \tmtexttt{mint} performs the
  folded integration. The grid is kept fixed at this stage. The array\\
  \tmtexttt{ \ \ \ \ integer ifold(ndimmax)\\
  \ \ \ \ \ common/cifold/ifold}\\
  controls the folding, i.e. \tmtexttt{ifold(k)} is the number of folds for
  \tmtexttt{x(k)}. The number of folds must be an integer divisor of the
  number of bins, which is fixed to 50. The array \tmtexttt{xgrid} and the
  value \tmtexttt{xint} must have already been filled by a previous call to
  \tmtexttt{mint} with \tmtexttt{ifl=0}. The upper bounding envelope of the
  folded function is also computed at this stage, and stored in the array
  \tmtexttt{ymax}.
\end{itemizedot}
The value and error of the integral are returned in \tmtexttt{ans} and
\tmtexttt{error}. Once \tmtexttt{ymax} has been setup in this way, one can
call \tmtexttt{gen} to generate events with the distribution of the folded
function. Notice that negative values of the called function are not allowed
at this stage. One calls\tmtexttt{\\
\ \ \ \ \ imode=0\\
\ \ \ \ \ call gen(fun,ndim,xgrid,ymax,imode,x)\\
\ \ \ \ \ imode=1\\
\ \ \ \ \ do j=1,10000\\
\ \ \ \ \ \ \ \ call gen(fun,ndim,xgrid,ymax,imode,x)\\
\ \ \ \ \ \ \ \ ...\\
\ \ \ \ \ enddo\\
\ \ \ \ \ imode=3\\
\ \ \ \ \ call gen(fun,ndim,xgrid,ymax,imode,x)}\\
where the call with \tmtexttt{imode=0} initializes the generation,
\tmtexttt{imode=1} generate the ndim-tuples, and \tmtexttt{imode=3} prints out
generation statistics. After a call to \tmtexttt{gen} with \tmtexttt{imode=1},
one can assume that the last call to the function \tmtexttt{fun} was performed
with the generated value of \tmtexttt{x}, so that parameters depending upon
\tmtexttt{x} that are stored in common blocks by \tmtexttt{fun} have
consistent values.

\end{document}